\pdfoutput=1
\documentclass[aps,prl,showpacs,preprintnumbers,twocolumn]{revtex4-1}

\usepackage[colorlinks=true, pdfstartview=FitV, linkcolor=blue, citecolor=blue, urlcolor=blue]{hyperref}
\usepackage[dvipdfmx]{graphicx}
\usepackage{amsmath}
\usepackage{color}

\begin{document}

\title{Berry phase in lattice QCD}

\author{Arata~Yamamoto}
\affiliation{Department of Physics, The University of Tokyo, Tokyo 113-0033, Japan}

\date{\today}
\begin{abstract}
We propose the lattice QCD calculation of the Berry phase which is defined by the ground state of a single fermion.
We perform the ground-state projection of a single-fermion propagator, construct the Berry link variable on a momentum-space lattice, and calculate the Berry phase.
As the first application, the first Chern number of the (2+1)-dimensional Wilson fermion is calculated by the Monte Carlo simulation.
\end{abstract}

\pacs{12.38.Gc, 11.15.Ha, 03.65.Vf}
\maketitle

\paragraph{Introduction.}
The Berry phase emerges in a variety of situations in physics \cite{Berry:1984}.
The most famous manifestation is the Aharanov-Bohm effect \cite{Aharanov:1959}.
The Berry phase plays key roles on the quantum Hall effect and topological insulators in condensed matter physics \cite{Thouless:1982zz}, rotating nuclei in nuclear physics \cite{Nikam:1987}, and quark confinement in particle physics \cite{Kondo:1998nw}.
There are too many examples to cover all of them.

Among various kinds of the Berry phase, let us focus on the Berry phase in momentum space of a fermion.
The Berry phase of electrons is essential for the quantum Hall effect and topological insulators.
They are classified by topological order, i.e., topology of the Berry phase, which differs from conventional order of symmetry breaking \cite{Xiao:2010}.
The Berry phase of chiral fermions describes anomalous current effects, e.g., the chiral magnetic effect and the chiral vortical effect \cite{Son:2012wh}.
Through the anomalous current effects, quantum anomaly and vacuum topology can be experimentally detected in heavy-ion collisions \cite{Kharzeev:2015znc} and in condensed matter materials \cite{Li:2014bha}.
A remarkable and common point is that these phenomena are protected by topology and exhibit dissipationless transport.
This opens up possibilities for phenomenological impact and technological application.

In simple quantum mechanical problems, the Berry phase can be derived analytically.
In realistic materials, however, analytical derivation is impossible and exact calculation relies on computational approaches \cite{Gradhand:2012}.
In quantum chromodynamics (QCD), there has been no full quantum calculation including interaction effects.
In this Letter, we would like to propose the first numerical calculation to compute the Berry phase in lattice QCD.
Since the lattice QCD simulation is the first-principle quantum calculation, this attempt enables the exact investigation of the Berry phase and topological order in QCD.

We focus on the Berry phase in momentum space of a single fermion.
In general, the Berry phase can be defined not only in momentum space but also in any other parameter space.
It would be possible to formulate similar numerical schemes for the Berry phase of other parameters.
Actually, the Berry phases of internal parameters were studied in the quantum Monte Carlo simulations of condensed matter systems \cite{Motoyama:2013}.

\paragraph{Formalism.}

Let us recall the conventional definition of the Berry phase in the continuum.
We consider the ground state $\tilde{\phi}(p)$ with the $d$-dimensional momentum $\vec{p}$.
Although the same argument holds for any states, we only consider the non-degenerate ground state for simplicity.
The Berry connection is defined by 
\begin{equation}
\label{eqA}
\tilde{A}_\mu (p) = -i \tilde{\phi}^\dagger(p) \frac{\partial}{\partial p^\mu} \tilde{\phi}(p)
,
\end{equation}
and the Berry curvature is
\begin{equation}
\label{eqF}
 \tilde{F}_{\mu\nu}(p) = \frac{\partial}{\partial p^\mu} \tilde{A}_\nu (p) - \frac{\partial}{\partial p^\nu} \tilde{A}_\mu (p)
.
\end{equation}
Intuitively, $\tilde{A}_\mu (p)$ and $\tilde{F}_{\mu\nu}(p)$ are the momentum-space counterpart of the coordinate-space gauge field $A_\mu(x)$ and the field strength $F_{\mu\nu} (x)$, respectively.
The Berry phase is defined by the line integral of the Berry connection along a closed path, or, equivalently, the surface integral of the Berry curvature
\begin{equation}
\gamma =
\oint dp^\mu \tilde{A}_\mu (p) = \int d S^{\mu\nu} \tilde{F}_{\mu\nu} (p)
.
\end{equation}

We introduce the above concept to lattice QCD.
In lattice QCD, the ($d$+1)-dimensional Euclidean path integral
\begin{equation}
Z = \int {\mathcal D}U \, {\rm det}D \, e^{-S} 
\end{equation}
is evaluated by the Monte Carlo method.
Firstly, we numerically generate the background gluon configurations  by the standard Monte Carlo sampling.
Then, we execute the following three steps for each gluon configuration.

(I) Ground-state projection.
Although the ground state can be obtained by the diagonalization of Hamiltonian, which is familiar in nonrelativistic theory, it is not familiar in lattice QCD.
Instead, the ground state is obtained by taking the long imaginary-time limit of a propagator.
The single-fermion propagator $D^{-1}(x,x')$ is given by the inverse of the Dirac operator $D(x,x')$.
A single-fermion state with a fixed momentum is given by
\begin{equation}
\label{eqproj}
 \tilde{\phi}(p)
= \sum_{\vec{x},\vec{x}'} e^{i \vec{p}\cdot (\vec{x}-\vec{x}')} D^{-1}(x,x') \phi_{\rm init}
\end{equation}
with $x=(\vec{x},\tau)$ and $x'=(\vec{x}',0)$.
Here, translational invariance is assumed and the center-of-mass motion is dropped.
When $\tau$ is taken large enough, excited-state components are suppressed and $\tilde{\phi}(p)$ is dominated by the ground-state component.
If the ground-state dominance is fulfilled, $\tilde{\phi}(p)$ depends only on $\vec{p}$ and is independent of $\tau$ up to a normalization constant.
The ground-state dominance is identified by the existence of a plateau in large $\tau$ region (as we will see later).
The initial state $\phi_{\rm init}$ can be chosen arbitrarily as long as it has nonzero overlap with the ground state.
This is the standard ground-state projection method used in lattice QCD, e.g., in the calculation of a hadron mass.

(II) Construction of the Berry link variable.
Discrete and finite coordinate space is mapped onto discrete and finite momentum space by the Fourier transformation.
The momentum space has the finite size, the so-called Brillouin zone, ($-\pi/a,\pi/a$] and the nonzero lattice spacing
\begin{equation}
\tilde{a} = \frac{2\pi}{La}
,
\end{equation}
where $L$ is the number of lattice site in one dimension.
The Berry phase is described by $d$-dimensional lattice gauge theory on this momentum-space lattice \cite{Fukui:2005wr}.
In lattice gauge theory, the connection is replaced by the link variable, which is an element of U(1) group.
We construct the Berry link variable
\begin{equation}
\label{eqU}
 \tilde{U}_\mu(p) = e^{i \tilde{a} \tilde{A}_\mu (p)} = \frac{ \tilde{\phi}^\dagger(p) \tilde{\phi}(p + \tilde{\mu}) }{| \tilde{\phi}^\dagger(p) \tilde{\phi}(p + \tilde{\mu}) |}
,
\end{equation}
where $\tilde{\mu}$ is the unit lattice vector in the $p^\mu$ direction.
We can show that Eq.~\eqref{eqU} reproduces Eq.~\eqref{eqA} in the continuum momentum limit $\tilde{a} \to 0$, i.e., the infinite spatial size limit $L \to \infty$.

(III) Calculation of the Berry phase.
In principle, the Berry phase is obtained by the phase of the Wilson loop, which is a product of $\tilde{U}(p)$, along a closed path.
However, if the loop is too large, the admissibility condition is violated and topology is destroyed \cite{Luscher:1998du}.
A better way is to calculate the Berry plaquette
\begin{equation}
\label{eqP}
\begin{split}
 \tilde{P}_{\mu\nu}(p) 
&= e^{i\tilde{a}^2\tilde{F}_{\mu\nu}(p)} 
\\
&= \tilde{U}_\mu(p) \tilde{U}_\nu(p+\tilde{\mu}) \tilde{U}^\dagger_\mu(p+\tilde{\nu}) \tilde{U}^\dagger_\nu(p)
,
\end{split}
\end{equation}
the Berry curvature
\begin{equation}
 \tilde{F}_{\mu\nu}(p) = {\rm Im} \ln \tilde{P}_{\mu\nu}(p)
,
\end{equation}
and the Berry phase
\begin{equation}
 \gamma = \sum_p \tilde{F}_{\mu\nu}(p)
\end{equation}
on the corresponding surface.

We repeat these three steps for all the gluon configurations, and then take the ensemble average
\begin{equation}
\langle O \rangle 
= \frac{1}{Z} \int {\mathcal D}U \, O \, {\rm det}D \, e^{-S}
= \frac{1}{N_{\rm conf}} \sum_{\{U\}} O
.
\end{equation}
The observable $O$ is, for example, the Berry phase $\gamma$.
Any observable which is a functional of the Berry link variable \eqref{eqU} is calculable.
This scheme is applicable to exact calculation of free fermions as well as the Monte Carlo simulation of interacting fermions.
For free fermions, the free Dirac operator is used and the ensemble average is not taken.

We note the local gauge invariance of this framework.
There are two gauge fields: the U(1) Berry field in momentum space and the SU(3) gluon field in coordinate space.
The U(1) gauge invariance in momentum space is preserved when the observable is given by loops in momentum space.
For example, the U(1) gauge fixing is not necessary to calculate $\tilde{P}_{\mu\nu}(p)$ (or $\tilde{F}_{\mu\nu}(p)$) but necessary to calculate $\tilde{U}_\mu(p)$ (or $\tilde{A}_\mu (p)$).
On the other hand, the SU(3) gauge invariance in coordinate space is explicitly broken because the momentum of a single fermion is gauge dependent.
This can be seen in Eq.~\eqref{eqproj}, where the Fourier transformation of the single-fermion propagator $D^{-1}$ is gauge dependent.
Thus, we need some SU(3) gauge fixing for each gluon configuration; otherwise, the expectation values of observables are zero because of the Elitzur theorem \cite{Elitzur:1975im}.
For SU(3) gauge fixing, any gauge choice is possible as long as all spatial gauge degrees of freedom are fixed.

\paragraph{Numerical results.}

Let us apply the above framework to a simple example.
We consider the (2+1)-dimensional Wilson-Dirac operator
\begin{equation}
\begin{split}
\label{eqD}
 D(x,x') 
=& \ (ma+3) \delta_{x,x'}
\\
&- \frac{1}{2} \sum_{\mu=1}^{3} \bigg[ \left(1-\sigma_\mu \right) U_\mu(x) \delta_{x+\hat{\mu},x'}
\\
&+ \left(1+\sigma_\mu \right) U_\mu^\dagger(x') \delta_{x-\hat{\mu},x'} \bigg]
,
\end{split}
\end{equation}
where $a$ is the lattice spacing in coordinate space and $\hat{\mu}$ is the unit lattice vector in the $x^\mu$ direction.
$U_\mu(x)$ is the SU(3) link variable, i.e., the lattice gluon field.
It is known that the momentum-space topology of the Wilson fermion is trivial for $m>0$ and nontrivial for $m<0$ \cite{Qi:2008ew}.
The (2+1)-dimensional spinor has two states, one ground state (negative energy $p_\tau<0$) and one excited state (positive energy $p_\tau>0$), which have the same energy but opposite signs.
Since the Fourier transformation of Eq.~\eqref{eqD} satisfies the relation
\begin{equation}
\sigma_2 \tilde{D}(p_x,p_y,p_\tau) \sigma_2 = \tilde{D}(-p_x,p_y,-p_\tau),
\end{equation}
the opposite energy states have the opposite Berry curvatures. 
This property holds even for interacting cases.

\begin{figure}[h]
 \includegraphics[width=.49\textwidth]{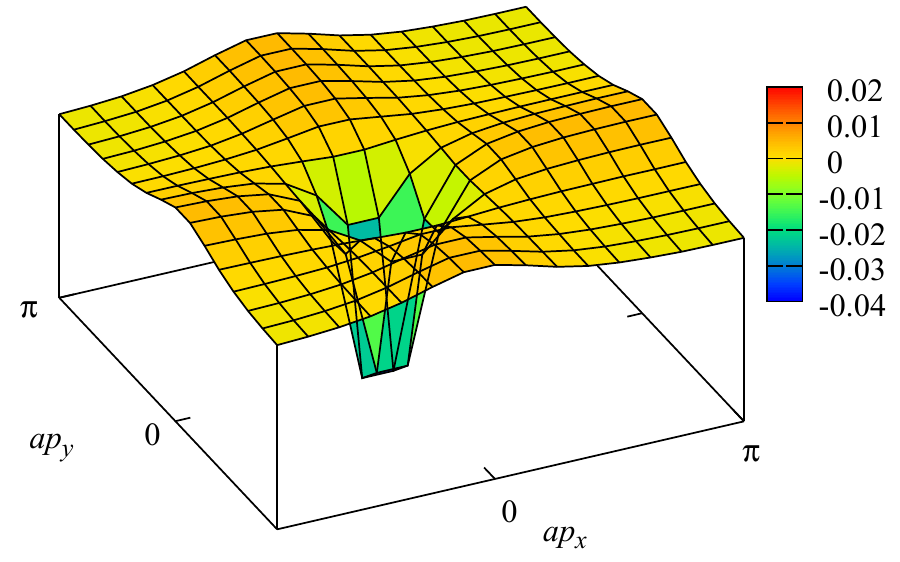}
 \caption{\label{fig1} 
Berry curvature $\tilde{a}^2\tilde{F}_{xy}(p)$ for $ma = 0.5$, which corresponds to $N=0$.
The data of free theory at $\tau/a = 12$ are shown.
 }
\end{figure}

\begin{figure}[h]
 \includegraphics[width=.49\textwidth]{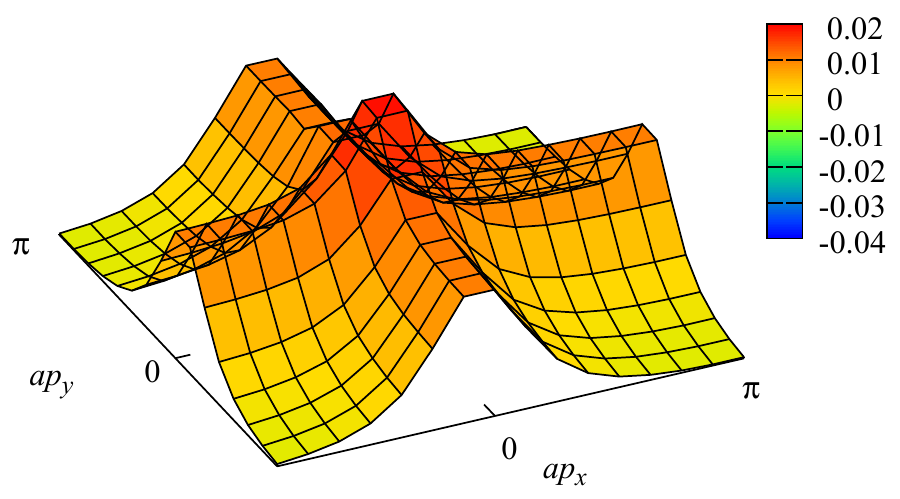}
 \caption{\label{fig2} 
Berry curvature $\tilde{a}^2\tilde{F}_{xy}(p)$ for $ma = -0.5$, which corresponds to $N=1$.
The data of free theory at $\tau/a = 12$ are shown.
 }
\end{figure}

\begin{figure}[h]
 \includegraphics[width=.49\textwidth]{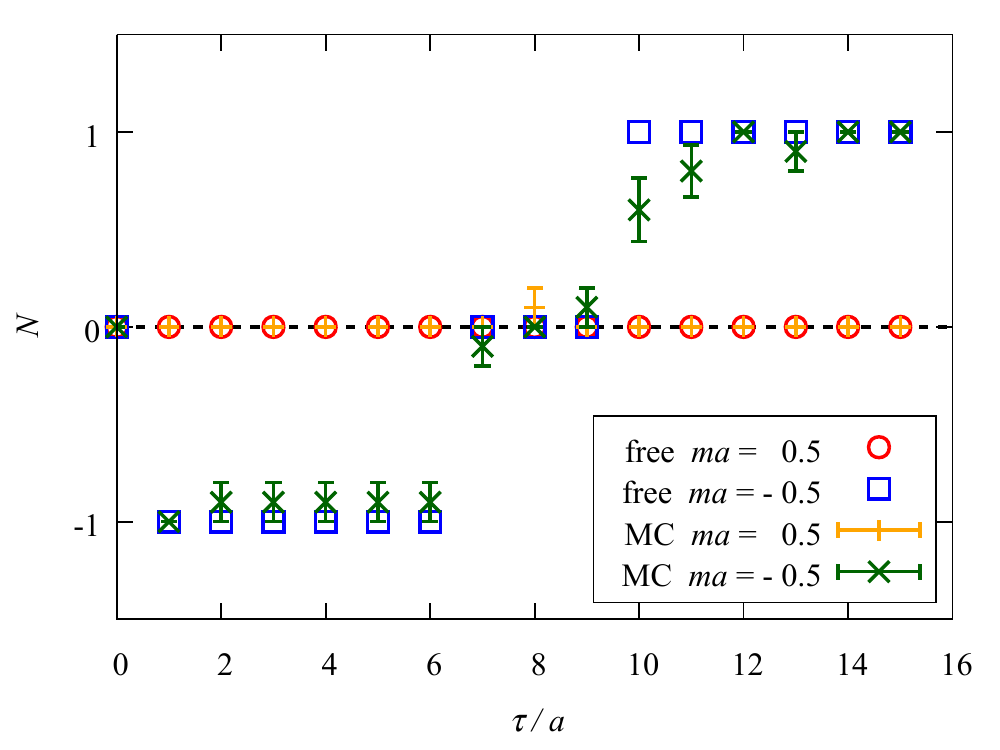}
 \caption{\label{fig3} 
First Chern number $N$ as a function of imaginary time $\tau$.
The data of free theory and the Monte Carlo (MC) simulation are shown.
 }
\end{figure}

First we analyze non-interacting cases $U_\mu(x)=1$.
The lattice volume is $L_xL_yL_\tau = 16^3$.
The boundary conditions are taken to be periodic in the $x$ and $y$ directions and antiperiodic in $\tau$ direction.
The initial state is $\phi_{\rm init} = (1,1)^\top$.
In Figs.~\ref{fig1} and \ref{fig2}, we show the Berry curvature $\tilde{F}_{xy}(p)$ of the free Wilson fermion with $ma=0.5$ and $-0.5$, respectively.
Both in Figs.~\ref{fig1} and \ref{fig2}, we see the peak of the Berry curvature at $p_x=p_y=0$.
However, nonzero curvature does not necessarily mean nontrivial topology.
The topological order parameter is not the local Berry curvature but the integral of the Berry curvature in whole momentum space
\begin{equation}
N = \frac{1}{2\pi} \sum_p \tilde{F}_{xy}(p)
.
\end{equation}
This quantity is known as the first Chern number, which explains the quantum Hall effect.
It must be integer-valued.
In Fig.~\ref{fig3}, we plot the first Chern number $N$ as a function of imaginary time $\tau$.
For $m=0.5$, the result is independent of $\tau$ and thus the Chern number is trivially $N=0$.
For $m=-0.5$, the result depends on $\tau$.
For the Chern number of the ground state, we look at the value in large $\tau$ region.
From the plateau in $\tau = 10$-15, we conclude that the Chern number is $N=1$ for $m=-0.5$.

Summarizing the results of the calculations, we obtained
\begin{equation}
 N = 
\begin{cases}
    0 & (ma>0) \\
    1 & (0>ma>-2) \\
    -2 & (-2>ma>-4) \\
    1 & (-4>ma>-6) \\
    0 & (-6>ma) \\
\end{cases}
.
\end{equation}
The Chern number changes when gapless modes appear.
The change of the Chern number is $\delta N(ma=0)=1$, $\delta N(ma=-2)=-3$, $\delta N(ma=-4)=3$, and $\delta N(ma=0)=-1$.
These values are consistent with the numbers of chiral modes of the Wilson fermion.
Note that the result is different from the non-relativistic Wilson-Dirac model \cite{Qi:2008ew} because the Wilson term in the $\tau$ direction in Eq.~\eqref{eqD} gives additional gapless modes.

Next we study interacting cases in the Monte Carlo simulation.
The SU(3) link variable $U_\mu(x)$ is generated by the quenched Monte Carlo simulation of the (2+1)-dimensional Wilson gauge action.
Although fermion loops are neglected in the quenched approximation, they are inessential for the calculation of the Chern number, and thus the quenched approximation is expected to be valid for this first qualitative application.
The lattice coupling is $\beta = 6/g^2 = 22$ and other parameters are the same as in the non-interacting case.
The initial state is $\phi_{\rm init} = (1,1)^\top \, ({\rm spinor})\otimes (1,1,1)^\top \, ({\rm color})$.
The two-dimensional Coulomb gauge fixing is numerically performed for the SU(3) link variable.
We take the ensemble average
\begin{equation}
N = \left\langle \frac{1}{2\pi} \sum_p \tilde{F}_{xy}(p) \right\rangle
.
\end{equation}
As shown in Fig.~\ref{fig3}, the free result and the Monte Carlo data give the same Chern number.
This is consistent with the fact that the Chern number is robust against interaction as long as the gap structure is topologically unchanged \cite{Niu:1985}.
We numerically checked that the result is also robust against the change of simulation parameters, such as the lattice coupling (i.e., discretization) and the lattice size (i.e., finite volume and nonzero temperature).

We comment on one technical question about Fig.~\ref{fig3}.
What is the plateau of $N=-1$ in $\tau = 1$-6?
The answer is the excited state.
Although the ground-state projection is the conventional one in the hadron mass calculation, there are a few differences.
The single-fermion propagator is not invariant under time reversal.
The lowest energy state is enhanced in positive $\tau$ direction and the highest energy state is enhanced in negative $\tau$ direction.
Because of (anti)periodicity, the lowest energy state is dominant in large $\tau$ region and the highest energy state is dominant in small $\tau$ region .
Therefore the plateau of the highest-energy excited state appears in $\tau = 1$-6.
As explained above, the Chern number of the highest energy state is opposite to that of the lowest energy state, and thus $N=-1$.

\paragraph{Future perspective.}

We formulated the lattice QCD calculation for the Berry phase.
The most important point is that this calculation exactly includes interaction and quantum effects.
We saw that it works well for the first Chern number of the (2+1)-dimensional Wilson-Dirac operator.
Since the application to higher dimensions is straightforward, we can study the Berry physics of real QCD in 3+1 dimensions.
We can also study the (4+1)-dimensional domain-wall fermion, which has the second Chern number, from the viewpoint of topological order  \cite{Kimura:2015ixh}.
The application to condensed matter systems is also possible.
For instance, we can study topological orders of two-dimensional and three-dimensional electron systems, e.g., graphenes and Dirac semimetals, with interaction.

In general, topology of the Berry phase is robust against small perturbations if the energy gap between the ground state and the excited states is large.
This robustness validates analytical calculation with noninteracting approximation.
However, it can be contaminated by nonperturbative interactions.
The presence of external fields or crystalline structures makes the calculation even more difficult.
Nonperturbative lattice simulation is necessary to treat all such things.

Since the concept of the Berry phase is quite general, there are many extensions of this framework:
The Berry phase can be defined for any state other than the ground state.
In lattice QCD, a low-lying excited state is obtained by the excited-state projection method, which is known in the hadron mass calculation \cite{Gattringer:2007da}.
When eigenstates have degeneracy, the Berry phase becomes non-Abelian \cite{Wilczek:1984dh}.
The Berry link variable becomes non-Abelian and non-Abelian topology appears on a momentum-space lattice.
In addition, it would be possible to formulate the lattice calculation for the Berry phase of other momenta or other internal parameters.

\begin{acknowledgments}
The author was supported by JSPS KAKENHI Grant Number 15K17624. 
The numerical simulations were carried out on SX-ACE in Osaka University.
\end{acknowledgments}

\end{document}